\documentclass[notitlepage,floatfix,aps,twocolumn,amsmath,amssymb,superscriptaddress,10pt]{revtex4-2}
\usepackage[pretty,uselistings,final]{revquantum}

\usepackage{comment}

\usepackage{xcolor}
\usepackage{soul}
\usepackage{graphicx}

\definecolor{britishracinggreen}{rgb}{0.0, 0.26, 0.15}
\definecolor{forestgreen}{rgb}{0.13, 0.55, 0.13}
\definecolor{applegreen}{rgb}{0.55, 0.71, 0.0}
\definecolor{byzantine}{rgb}{0.74, 0.2, 0.64}
\definecolor{lapislazuli}{rgb}{0.15, 0.38, 0.61}
\definecolor{chocolate(web)}{rgb}{0.82, 0.41, 0.12}

\begin{document}
\title{Frequency-bin entanglement from domain-engineered down-conversion
}

\author{Christopher L. Morrison}
\email[Correspondence: ]{chrislmorrison93@outlook.com}
\affiliation{Institute of Photonics and Quantum Sciences, School of Engineering and Physical Sciences, Heriot-Watt University, Edinburgh EH14 4AS, UK}
\author{Francesco Graffitti}
\affiliation{Institute of Photonics and Quantum Sciences, School of Engineering and Physical Sciences, Heriot-Watt University, Edinburgh EH14 4AS, UK}
\author{Peter Barrow}
\affiliation{Institute of Photonics and Quantum Sciences, School of Engineering and Physical Sciences, Heriot-Watt University, Edinburgh EH14 4AS, UK}
\author{Alexander Pickston}
\affiliation{Institute of Photonics and Quantum Sciences, School of Engineering and Physical Sciences, Heriot-Watt University, Edinburgh EH14 4AS, UK}
\author{Joseph Ho}
\affiliation{Institute of Photonics and Quantum Sciences, School of Engineering and Physical Sciences, Heriot-Watt University, Edinburgh EH14 4AS, UK}
\author{Alessandro Fedrizzi}
\affiliation{Institute of Photonics and Quantum Sciences, School of Engineering and Physical Sciences, Heriot-Watt University, Edinburgh EH14 4AS, UK}

\begin{abstract}

Frequency encoding is quickly becoming an attractive prospect for quantum information protocols owing to larger Hilbert spaces and increased resilience to noise compared to other photonic degrees of freedom. To fully make use of frequency encoding as a practical paradigm for QIP, an efficient and simple source of frequency entanglement is required. Here we present a single-pass source of discrete frequency-bin entanglement which does not use filtering or a resonant cavity. We use a domain-engineered nonlinear crystal to generate an eight-mode frequency-bin entangled source at telecommunication wavelengths. Our approach leverages the high heralding efficient and simplicity associated with bulk crystal sources.
\end{abstract}

\date{\today}

\maketitle
\begin{figure}[b!]
    \begin{center}
    \includegraphics[width=0.9\columnwidth]{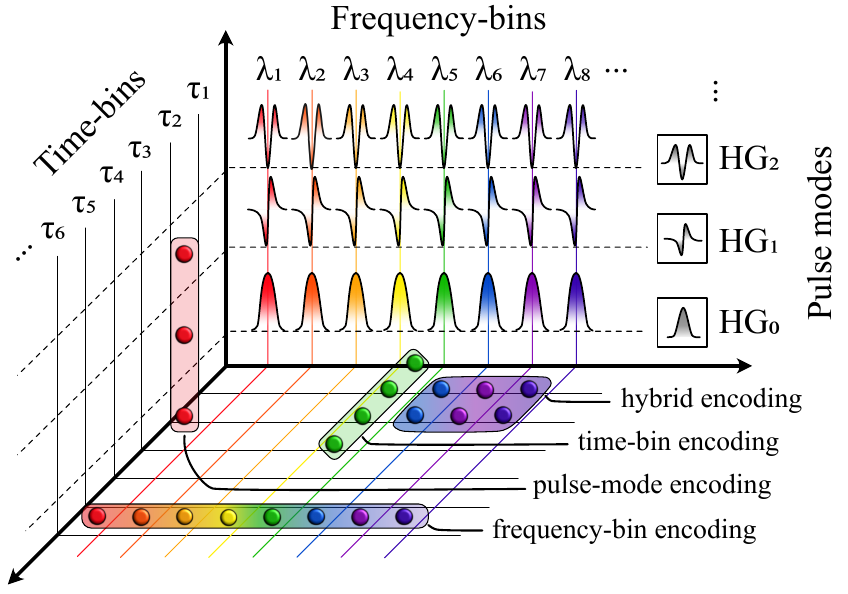}
    \caption{Encoding schemes in the time-frequency domain. Time-bin encoded states shown in green, encode bits in the arrival time of the photon $\tau_i$. Frequency-bin encoded states use the colour of the photons to encode information, $\lambda_i$. Pulse-mode encoding uses the longitudinal modes of single photons, here represented by Hermite-Gaussian modes, to encode information.
    Provided the time and frequency scales are well separated \cite{reimer_high-dimensional_2019} information can be encoded in a hybrid manner using combinations of time-bins, frequency-bins and pulse modes, shown here with two time-bins and three frequency-bins (purple rectangle).}
    \label{fig:concept}
    \end{center}
\end{figure}

Quantum photonics is increasingly exploiting the time-frequency degree of freedom (DoF) due to its compatibility with telecom infrastructure, the potential for greater information capacity per photon and improved resilience to noise~\cite{ding_high-dimensional_2017,PhysRevX.9.041042}. Time-frequency encodings can be broadly divided into three categories: time-bin encoded, frequency-bin encoded, and intensity-nonorthogonal ``pulse-mode'' encoded states (see Fig.~\ref{fig:concept}).

Time-bin states are encoded in discrete arrival times at the detection apparatus with a time separation exceeding the duration of the time-bins. Time-bin experiments date back  several decades, the encoding is easily manipulated with unbalanced interferometers and phase shifters and detected using standard single-photon detectors \cite{PhysRevLett.82.2594,PhysRevA.66.062308}. The complement of time-bin encoding is frequency-bin encoding, with photons localised in non-overlapping spectral regions \cite{PhysRevLett.103.253601,PhysRevA.82.013804}. Frequency-bin states can be manipulated and detected using electro-optic modulators \cite{PhysRevA.89.052323} and Fourier transform pulse shapers, which can in principle be lossless \cite{PhysRevLett.125.120503}. This has allowed for full tomography of frequency-bin entangled photon pairs \cite{kues_-chip_2017,lu2021quantum} and generation of on-chip cluster states \cite{reimer_high-dimensional_2019}.
Pulse-mode encoded states are the most recent addition to the experimental time-frequency toolbox, having initially been studied in the context of spectral entanglement in pulsed parametric down-conversion (PDC) \cite{law_continuous_2000}. Pulse-modes are overlapping in both spectral and temporal amplitude, their orthogonality is maintained by considering the temporal or spectral phase between different states. Due to this intensity nonorthogonality pulse-modes offer a way to extend time-frequency encoding beyond purely time or frequency-bin states (see Fig. \ref{fig:concept}). The introduction of the quantum pulse gate \cite{Eckstein:11} has made it possible to arbitrarily manipulate and measure pulse-modes. 
In this work we focus on the generation of frequency-bin encoded states. 

\begin{figure*}
\begin{center}
\includegraphics[width=.85\textwidth]{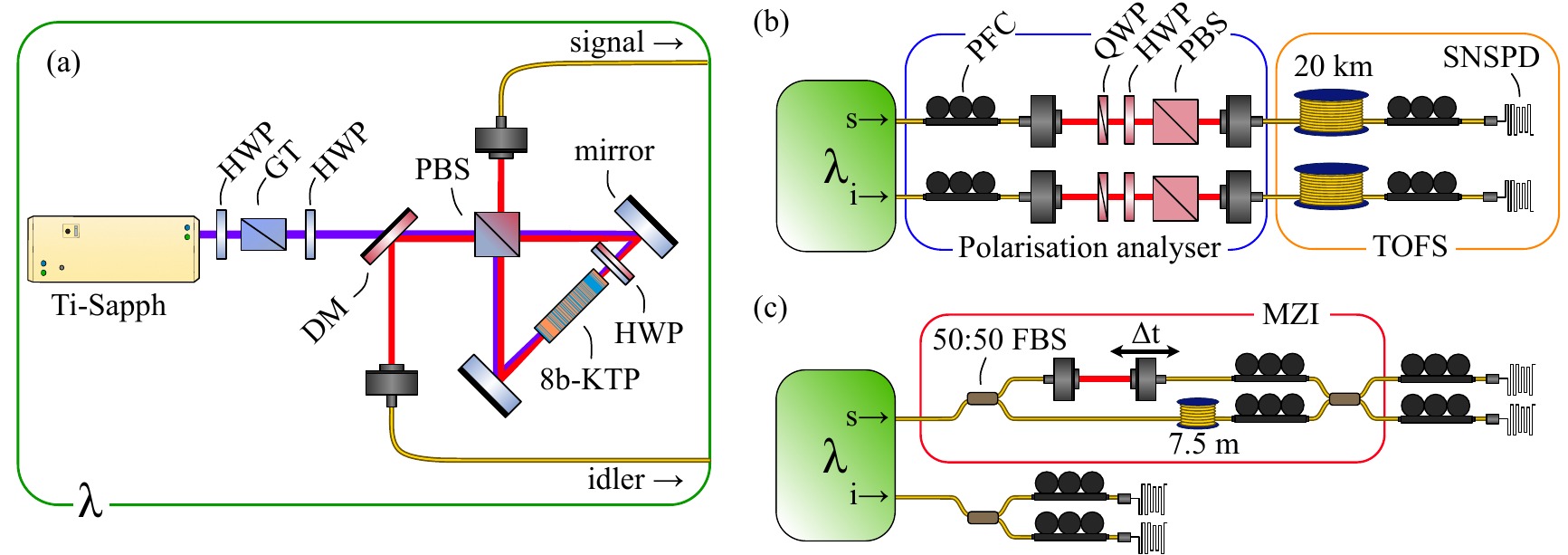}
\caption{Experimental setup. \textbf{(a)} An 80~MHz, 1.3~ps Ti-Sapphire laser is sent through a half wave-plate (HWP) and Glan-Taylor polariser (GT) for intensity and polarization control. The 8b-KTP is embedded in a Sagnac interferometer which allows the generation of polarisation entanglement. 
Down-converted photon pairs (signal and idler) are separated by a polarising beam splitter (PBS) then coupled into single mode fibres and sent to characterisation stages. 
\textbf{(b)} A polarisation-resolved time-of-flight spectrometer consisting of a standard polarisation tomography setup with spools of optical fibre for joint measurements on polarisation and frequency. The polarisation analyser consists of in-fibre polarisation controllers (PFC), a quarter-wave plate (QWP), half-wave plate (HWP) and PBS project the photons onto arbitrary polarisation states before they enter the time-of-flight spectrometer (TOFS). The TOFS is based on 20~km of spooled optical fibre, whose dispersion maps the frequency of the photon pairs onto arrival time at the superconducting nanowire single photon detectors (SNSPDs). The photon time-of-arrival is recorded using a fast (1~ps bin width) time tagger. 
\textbf{(c)} Heralded two-photon Hong-Ou-Mandel interferometer. The signal photon is sent through a 50:50 fibre beamsplitter (FBS) which probabilistically directs each photon into an unbalanced Mach-Zehnder interferometer (MZI). One MZI path contains a $\sim$7.5~m fibre which introduces a temporal delay corresponding to three laser clock cycles. The second MZI path contains an adjustable free-space delay, $\Delta t$. Two signal photons passing through the MZI then interfere on a second 50:50 FBS before detection with SNSPDs. The idler mode is sent to a 50:50 FBS then detected with SNSPDs to herald successive probabilistic emissions from the source. Coincidence counts are recorded for successful four-fold events. 
}
\label{experimental layout}
\end{center}
\end{figure*}

Discrete frequency-bin entangled states have been generated by filtering broadband correlated biphotons~\cite{chang_648_2021,Lingaraju:19} which is attractive due to the compatibility with commercial wavelength-division multiplexing (WDM) telecom components.
Frequency-bin entanglement can also be generated on-chip by using microresonators which produce photon pairs across multiple cavity resonances~\cite{kues_-chip_2017}---an intrinsically stable approach which can be scaled to a large number of sources on one chip. The main drawback of both of these approaches is optical loss due to filtering and/or resonant losses inside the microresonators which drastically limits the achievable heralding efficiency. 
Techniques from ultrafast pulse shaping can also be used to generate time-frequency encoded states~\cite{Ansari:18} but arguably a simpler technique is to employ PDC with domain-engineered nonlinear crystals, which allows for the  tailoring of almost arbitrary bi-photon phasematching functions (PMF) ~\cite{Graffitti_2017}. 
Domain-engineered crystals have been used to generate Gaussian biphoton states for pure heralded single photons~\cite{Graffitti:18,Pickston:21} and maximally entangled pulse-modes in two dimensions~\cite{FranPulseMode}.
Dual-poled crystals have also been used to generate frequency-bin entanglement over two modes \cite{Kaneda:19}.

In this work we demonstrate an eight-mode frequency-bin entangled source centered at telecom wavelengths. We use a single pass of a pump beam to generate the frequency-bin entanglementment without the need for filtering. This process is in principle lossless and allows for high heralding efficiencies of over 60\%, corrected for detector efficiency. This is in contrast to typical frequency-bin entangled sources which are inherently lossy due to filtering or the use of an optical cavity.
Typical frequency-bin sources require CW pumping, our approach uses pulsed excitation which can be scaled up for use in multiphoton protocols, demonstrated by heralded two photon interference measurements. 
We verify eight-mode frequency entanglement through a combination of joint spectral intensity (JSI) and two photon interference measurements. Using the crystal inside a Sagnac interferometer (see Fig~\ref{experimental layout}(a)) we also demonstrate polarisation-frequency hyperentanglement with frequency-resolved polarisation tomography.

Our domain-engineering algorithm~\cite{Graffitti_2017} tracks a target PDC amplitude along the crystal and chooses the domain width and orientation on a domain-by-domain basis to maximise the overlap to the desired target PMF. The target PMF in this work was a comb of eight separate Gaussian peaks (see supplementary \ref{supp: crystal design}). The crystal was manufactured by \emph{Raicol Ltd}.
With the symmetric group-velocity matching condition in type-II PDC in potassium titanyl phosphate (KTP), the eight Gaussian peaks in the PMF function result in a spectrally entangled state. 
Using the Schmidt decomposition the biphoton state can be written as 
\begin{equation}
    \ket{\psi} = \sum_i \sqrt{\lambda_i} A_i^{\dagger}B_i^{\dagger}\ket{00},
\end{equation}
where the $A_i^{\dagger}, B_i^{\dagger}$ are broadband mode operators and $\lambda_i$ are the Schmidt weights \cite{law_continuous_2000}. For a maximally entangled eight-mode state $\lambda_i = 1/8$. The Schmidt number $K = 1 / \sum_i \lambda_i^2$ defines the effective number of modes in the PDC process, for a $n$-mode maximally entangled state $K = n$.

In this experiment the eight-bin KTP crystal (8b-KTP) crystal is pumped by a Ti-Sapphire laser with a base repetition rate of 80~MHz, a pulse duration of 1.3~ps and central wavelength of 777.85~nm which leads to degenerate photons at 1555.7~nm.
The width of the frequency bins are chosen to maximise the overlap with an eight-mode maximally entangled state for the available pump pulse duration. The pump is focused into the crystal with a 40~cm focal length lens for a beam waist of approximately 77~$\mu$m.
The pump light is removed with a silicon filter and long pass filter at 1400~nm before the photons are collected in single mode fibres. The photons are detected with superconducting nanowire single-photon detectors (SNSPDs) with a nominal quantum efficiency of 80\% and 50~ps jitter. Time-tags are recorded using a \emph{Hydraharp 400} time-tagger with 1~ps bin width.

Full characterisation of frequency-bin entanglement requires projective measurements on an informationally complete set of spectral modes.
While one basis can easily be measured using optical filters, measurements in coherent bases typically requires electro-optic modulation~\cite{PhysRevLett.125.120503,davis_measuring_2020} or nonlinear frequency conversion~\cite{Eckstein:11}.
However, biphoton interference measurements can confirm the presence of spectral entanglement without resorting to these methods. Anti-bunching in biphoton interference is an indicator of antisymmetry in the biphoton wavefunction~\cite{Fedrizzi_2009}. 
With a separable polarisation state the anti-bunching, clearly visible at $\pm 1$~ps delay, can be attributed to spectral entanglement. We measured an interference visibility of 97.9(3)\% which indicates the quality of the spectral entanglement generated directly from the source.
The repetition rate of the Ti-Sapphire laser is temporally multiplexed to 160~MHz using a delay line for the biphoton interference measurements. This reduces higher-order emission from the source while keeping the two-fold rate constant \cite{Broome:11}.

\begin{figure}
\begin{center}
\includegraphics[width=1\columnwidth]{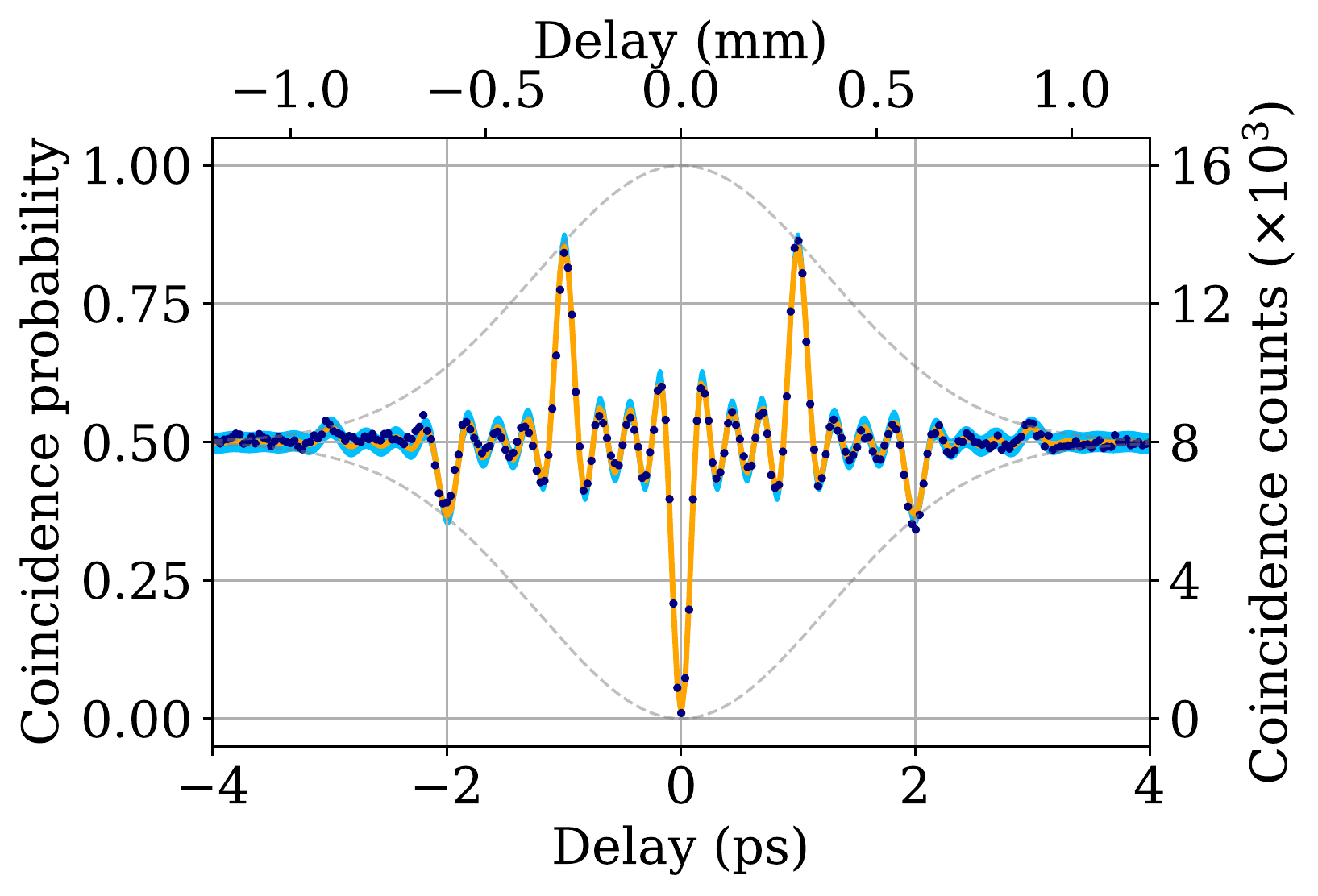}
\caption{Biphoton HOM interference data (blue dots) with line of best fit (orange) from theoretical model (supplementary \ref{two fold supp}) with bin width, bin separation and visibility as free parameters.  Coincidence counts are integrated over two seconds with an incident pump power of 400 mW. Counts are recorded every 0.1~mm using a motorised translation stage. The measured visibility is 97.9(3)\%.
The frequency-bin separation $\delta$ (see supplementary \ref{two fold supp}) extracted from the best fit parameters is found to be 499.7(3)~GHz which agrees with the designed spacing of 500~GHz.
The shaded blue region is the three sigma confidence region assuming Poissonian counting statistics. The grey dashed line shows the Gaussian envelope expected from a single frequency bin which bounds the visibility of the frequency beating.
}
\end{center}
\end{figure}

We measure the joint spectral intensity of the biphoton state using a dispersive fiber spectrometer, see Fig.~\ref{experimental layout}~(d). The fibre dispersion maps the frequency of the photons to arrival time at the detectors. This allows for a simple reconstruction of the joint spectral intensity using coincidence counts.
Reconstructing the joint spectral amplitude with full spectral phase information is possible~\cite{jizan_phase-sensitive_2016,davis_measuring_2020,triginer_understanding_2020} but typically requires well characterised reference beams or electro-optic effects. We probe possible spectral phase correlations using heralded two photon interference measurements and show the visibility is consistent with a flat spectral phase. 

\begin{figure*}
\centering
  \includegraphics[width = 1.9\columnwidth]{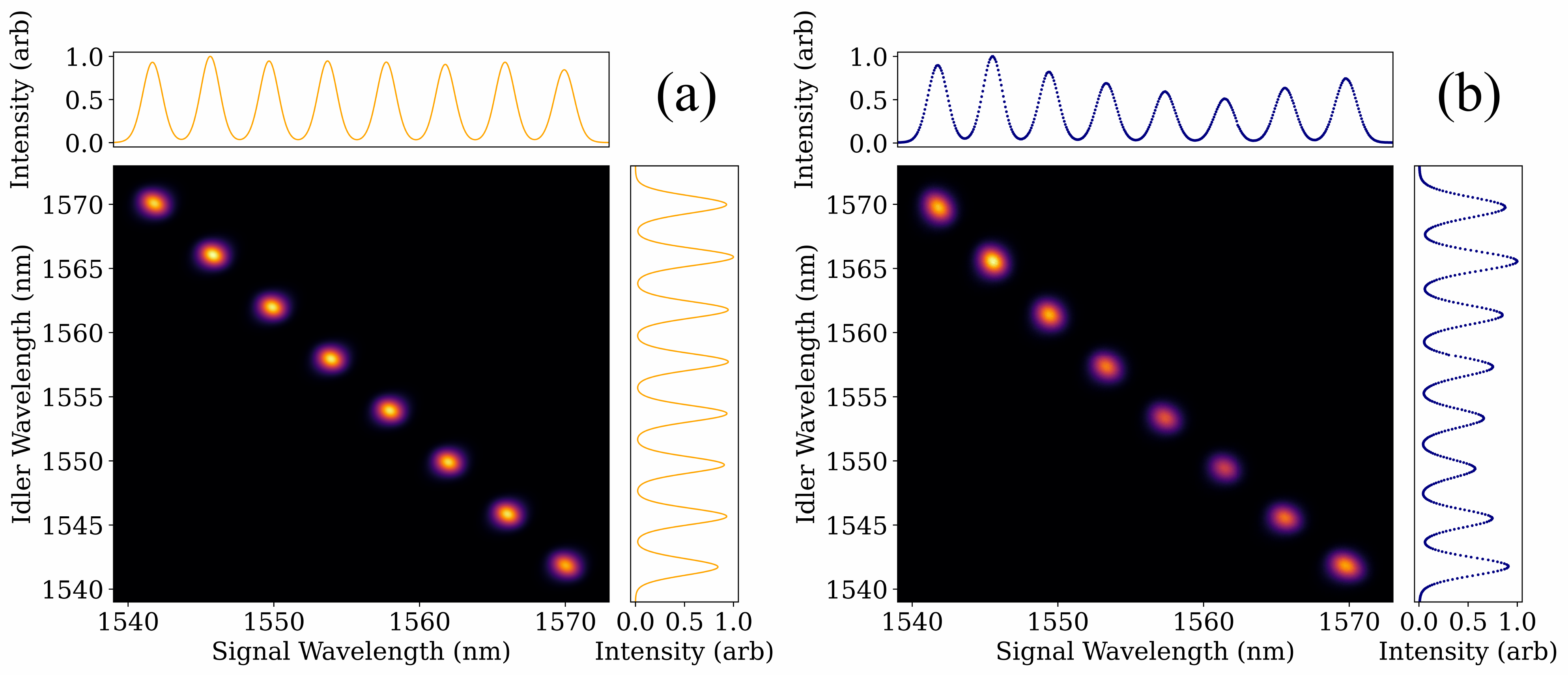}
\caption{Theoretical (a) and experimentally measured (b) joint spectral intensity. The JSI is reconstructed through TOFS on a 500x500 grid with each pixel corresponding to an arrival window of 25~ps, which corresponds to a spectral resolution of around 0.06~nm. The spectral range of the spectrometer is set by the temporal separation between pulses from the pump laser corresponding to 12.5~ns, which is wide enough to contain all eight frequency-bins. The number of frequency-bins could be extended at the cost of reducing the repetition rate of the pump laser. We include the marginal distributions on the left and top of the main JSI plot, with counts normalised to the maximum peak height. \label{JSI}}
\end{figure*}

We numerically calculate the Schmidt decomposition of the square root of the joint spectral intensity from $4.3\times10^7$ detected events.
We calculate an overlap of 96.01(1)\% to an eight-mode maximally entangled state by comparing Schmidt numbers to the ideal value of $\lambda_i = 1/8$ for the first eight Schmidt modes and zero otherwise. We also estimate a Schmidt number of $K= 7.018(3)$, both errors are quoted at three sigma estimated from 1000 rounds of Monte Carlo simulation assuming Poissonian counting statistics. The theoretical values based on the designed phasematching function are 98.5\% and 8.07 respectively. We attribute the difference in Schmidt number and fidelity to the unequal peak heights and non-zero extinction between the peaks, most clearly seen in the marginals of Fig.~\ref{JSI}. The average separation between peaks is measured to be 498~GHz which matches well with the designed separation of 500~GHz. The JSI is calibrated against marginal spectra measurements with a commercial single-photon spectrometer. 

In order to assess phase correlations in the joint spectrum we also carry out heralded two photon interference.
For an n-mode maximally entangled PDC process the heralded interference visibility is given by $1/n$, therefore we expect a visibility of 12.5\% for an eight-mode entangled state. We measure an interference visibility of 11.2(1.4)\% (see supplementary \ref{4foldsection}) which agrees with the expected visibility within experimental error. These measurements are consistent with a flat spectral phase across the joint spectrum.

Using the crystal in a Sagnac interferometer we generate a hyperentangled state in polarisation and frequency mode of the form 
\begin{equation}
    \ket{\psi} =\sum_{\substack{i=-4 \\i\neq 0}} ^4 \dfrac{1}{\sqrt{2}}\left(\ket{H}\ket{V} - e^{i \phi_i}\ket{V}\ket{H}\right)\otimes \dfrac{1}{\sqrt{\mathcal{N}_i}}  \ket{i}\ket{-i},
\end{equation}
with $\mathcal{N}_i$ related to the probability of emitting a pair of photons into the i-th frequency bin.
Due to the wavelength dependent retardance of the polarisation optics within the Sagnac source, each frequency-bin has a different phase $\phi_i$. 
We show that the polarisation entanglement persists across all eight bins by using TOFS after polarisation tomography. The TOFS measurement mimics the action of a WDM by temporally demultiplexing the frequency components into different arrival times.
We perform quantum state tomography using symmetric informationally complete (SIC) projections on the two-qubit polarisation state. This allows us to use fewer measurements when projecting with a single detector outcome, see supplementary \ref{freq pol tom} for more details.

The fidelity to the singlet state is calculated by applying a correction for the phases $\phi_i$ in post-processing. In a wavelength-division multiplexed scenario this phase can be corrected by appropriate waveplate settings. 
The average purity and fidelity are 88.7(3)\% and 92.6(1)\% respectively with a maximum (minimum) fidelity of 97.3\% (88.7\%), see supplementary \ref{freq pol tom}. Errors are calculated by 1000 rounds of Monte Carlo simulation assuming Poissonian counting statistics.
The drop in purities compared to other telecom wavelength Sagnac sources \cite{Jin:14} is attributed to the wavelength dependence of the polarisation optics inside the Sagnac and can be improved by using achromatic optics. The waveplates in the tomography setup also impart different unitaries across the full bandwidth of the downconverted photons which reduces the purity of the reconstructed state. This can be mitigated by reconstructing the polarisation state of each bin individually using WDM filters \cite{neumann2021experimental} or again using achromatic optics.

The efficient, pulsed source of frequency-bin entanglement presented in this work can be extended and improved in future work in multiple ways. 
Improving crystal fabrication to match lengths available with other materials such as lithium niobate would allow for denser packing of spectral features using domain engineering. With a longer crystal the frequency-bin spacing can be made to more closely match the ITU standard 100~GHz grid, which will allow for efficient integration with future WDM quantum networks \cite{wengerowsky_entanglement-based_2018,siddarthNetwork}.
As the frequency-bin entanglement is generated with pulsed excitation it would be possible to efficiently carry out Bell state measurements and therefore entanglement swapping in a WDM network. This would provide a way to connect users across initially unconnected WDM networks using a central node to carry out Bell state measurements.

The domain engineering technique used here could be combined with other methods of shaping the joint spectrum such as pump shaping to produce time-frequency grid states \cite{PhysRevA.102.012607} or using time-frequency synthesis techniques \cite{RBJOptSynth} to double the number of spectral features produced from domain engineering. 
An open question for future work is if domain engineering can be used to generate biphotons in hybrid encodings (see Fig.~\ref{fig:concept}) with a combination of pulse-modes, time-bins or frequency-bins, which could be used to directly generate cluster states in the time-frequency degree of freedom \cite{reimer_high-dimensional_2019}.

We thank B. D. Gerardot and M. Malik for loan of equipment. This
work was supported by the UK Engineering and Physical
Sciences Research Council (Grant No. EP/T001011/1).
FG acknowledges studentship funding from EPSRC
under Grant No. EP/L015110/1.

\bibliography{references.bib}

\clearpage

\onecolumngrid
\section{Supplementary Material}
\subsection{Crystal design}
\label{supp: crystal design}

The symmetric group velocity matching conditions in type-II potassium titanyl phosphate (KTP) at 1555~nm allow for spectrally entangled states by shaping the phasematching function $\phi\left(\Delta k\left(\omega_i,\omega_s\right)\right)$.
The phasematching function defines a nonlinearity profile along the crystal $g\left(z\right)$ by a Fourier transform.

\begin{equation}
    \phi\left(\Delta k\left(\omega_i,\omega_s\right)\right) = \int g\left(z\right)e^{i\Delta k\left(\omega_i,\omega_s\right)}\text{d}z
\end{equation}

For periodically poled crystals $g\left(z\right)$ is a box function defined over the crystal length but other nonlinearity profiles can be realised using domain engineering algorithms \cite{Graffitti_2017}.

For a frequency-bin entangled biphoton state the phasematching function is a series of Gaussian functions centered at different $\Delta k$ and therefore different $\omega_i,\omega_s$.

\begin{equation}
   \phi\left(\Delta k;\delta,\varsigma\right) = \sum_j ^n \exp\left(-\dfrac{\varsigma^2 \left(\Delta k -\Delta k_0 - \left( j+\frac{1}{2}\right)\delta\right)^2 }{2}\right)
   +  \exp\left(-\dfrac{\varsigma^2\left(\Delta k -\Delta k_0 + \left( j+\frac{1}{2}\right)\delta\right)^2  }{2}\right),
   \label{freq_bin_pmf}
 \end{equation}

where $\varsigma$ determines the width of the Gaussian peaks and $\delta$ determines the spacing between the peaks for $2n$ peaks.

For the 8 bin design (8b-KTP) used in this work the nonlinearity profile is given by the Fourier transform of equation \ref{freq_bin_pmf} for $n=4$

\begin{equation}
    g\left(z\right) = \dfrac{2}{\varsigma} \exp\left(i\Delta k_0 z -\frac{z^2}{2\varsigma^2}\right)\sum_{n=0} ^5 \cos\left(\left(2n+1\right)\frac{\delta}{2} z\right).
\end{equation}

\begin{figure}[H]
\begin{center}
\includegraphics[width=0.5\columnwidth]{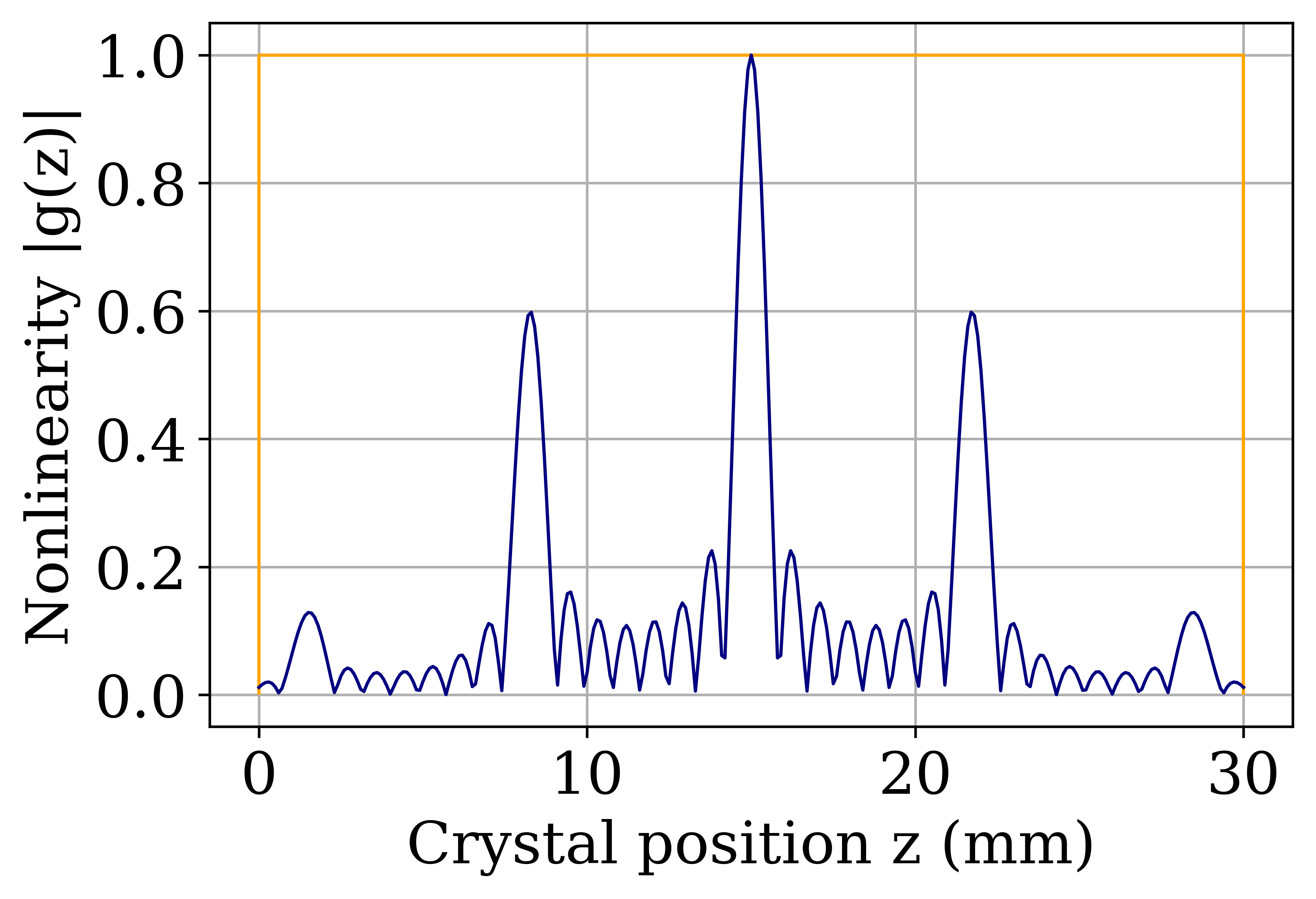}
\caption{Theoretical nonlinearity profile of the 8b-KTP crystal used in this experiment (navy), a periodically poled crystal is shown (orange) for comparison. The crystal length is 30 mm which matches the crystal used. With the overall reduced nonlinearity we measure a brightness of 450 coincidences/mW. }
\end{center}
\end{figure}

The parameters are chosen such that the spacing between the peaks is 500 GHz to match ITU 100 GHz grid for wavelength-division multiplexing. The peak width $\varsigma = L/4.5$ is set to maximise overlap with an eight-mode maximally entangled state. This value results in each frequency-bin being close to separable \cite{Pickston:21}, which could be used a pure heralded single-photon source with frequency resolved detection on the heralding arm. The theoretical purity based on the design parameters is 98.6\%.

\begin{figure*}[t]
\begin{center}
\includegraphics[width=1\columnwidth]{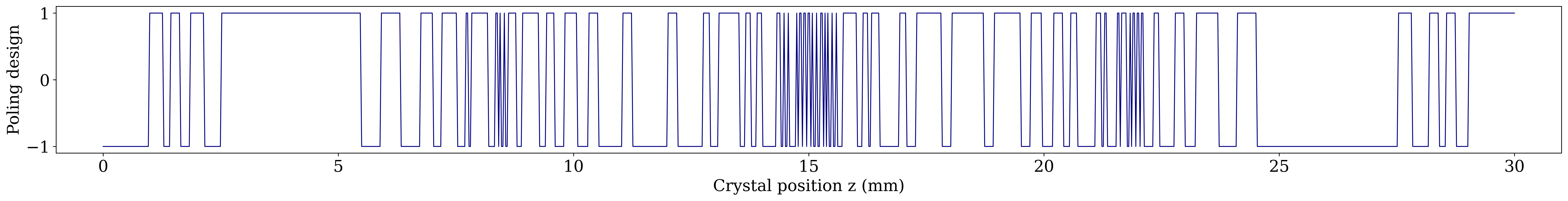}
\caption{The poling design for the 8b-KTP crystal.}
\end{center}
\end{figure*}

The domain widths are initially restricted to match a periodically poled crystal which is approximately 23~$\mu$m. The algorithm is then used with smaller domain widths of 11.5~$\mu$m corresponding to half the periodic poling domain width. For the parameters chosen in this work the smaller domain sizes did not improve the simulated overlap to an eight-mode maximally entangled state. This is expected due to the relatively long crystal length of 30~mm, sub-coherence domain engineering is expected to be relevant for crystals on the order of 1~mm. With larger domain sizes our algorithm works in a similar way to other domain-engineering algorithms \cite{Tambasco:16}.

\subsection{Two photon interference} \label{two fold supp}

To simplify the interference calculation we approximate the pump spectrum as a Gaussian function

\begin{equation}
    \alpha\left(\omega_s+\omega_i\right) = \e^{-\dfrac{\left(\omega_s+\omega_i\right)^2}{2 \sigma^2}}.
\end{equation}

The phasematching function is approximated as a sum of well separated Gaussian functions.

\begin{equation}
    \phi\left(\omega_i,\omega_s\right) = \sqrt{\dfrac{1}{\pi \sigma^2 \left(n+1\right)}}\sum\limits_{j=0}^n \left\{\e^{-\dfrac{\left(\omega_i-\omega_s+\delta\left(j+1/2\right)\right)^2}{2\sigma^2}} + \e^{-\dfrac{\left(\omega_i-\omega_s-\delta\left(j+1/2\right)\right)^2}{2\sigma^2}}\right\},
\end{equation}

where the number of bins is given by $2\left(n+1\right)$. The pump bandwidth and phasematching bandwidth are approximated as equal as the individual bins are close to separable as seen in the joint spectrum and in previous work \cite{Graffitti:18,Pickston:21}.

The coincidence probability for two photons from the same pulse is then calculated as

\begin{equation}
    p_{2}\left(\tau\right) =\dfrac{1}{2} - \dfrac{1}{2} \intop\text{d}\omega_i \intop\text{d}\omega_s  f^*\left(\omega_i,\omega_s\right)  f\left(\omega_s,\omega_i\right)e^{i\left(\omega_i-\omega_s\right)\tau},
\end{equation}

with $f\left(\omega_i,\omega_s\right)= \alpha\left(\omega_i+\omega_s\right)\phi\left(\omega_i,\omega_s\right)$. For well separated freuqency-bins, swapping the order of summation and integration in equation 3 and evaluating the integral over signal and idler frequencies give the coincidence probability as,

\begin{equation}
    p_{2}\left(\tau\right) = \dfrac{1}{2} - \dfrac{1}{2\left(n+1\right)} 
    \sum\limits_{j=0}^{n} \e^{-\dfrac{\left(\delta+2j\delta\right)^2 +\sigma^4\tau^2}{4\sigma^2}}\left(1+\e^{\dfrac{\left(\delta+2j\delta\right)^2}{4\sigma^2}}\cos\left(\dfrac{\delta \tau}{2}+j\delta\tau\right)\right).
\end{equation}

\subsection{Heralded two photon interference}
\label{4foldsection}

The coincidence probability for the heralded two photon interference is given by, 

\begin{equation}
    p_{4}\left(\tau\right) = \dfrac{1}{2} - \dfrac{1}{2}\intop\text{d}\omega_{i1}\intop\text{d}\omega_{s1}\intop\text{d}\omega_{i2}\intop\text{d}\omega_{s2}f^*\left(\omega_{i1},\omega_{s2}\right)f^*\left(\omega_{i2},\omega_{s1}\right)f\left(\omega_{i1},\omega_{s1}\right)f\left(\omega_{i2},\omega_{s2}\right)\e^{i\left(\omega_{s1} - \omega_{s2}\right)\tau}.
\end{equation}

Again swapping the order of summation and integration and evaluating the integral over signal and idler frequencies give the coincidence probability as,

\begin{equation}
    p_{4}\left(\tau\right) = \dfrac{1}{2} - \dfrac{1}{4\left(n+1\right)^2}\sum\limits_{j=0}^{n}\e^{-\dfrac{\left(\delta+2j\delta\right)^2 +\sigma^4\tau^2}{4\sigma^2}}\left(1+\e^{\dfrac{\left(\delta+2j\delta\right)^2}{4\sigma^2}} +2\cos\left(\frac{1}{4} \left(\delta+2j\delta\right)\tau\right) \right)
\end{equation}

\begin{figure}[ht]
\begin{center}
\includegraphics[width=0.5\columnwidth]{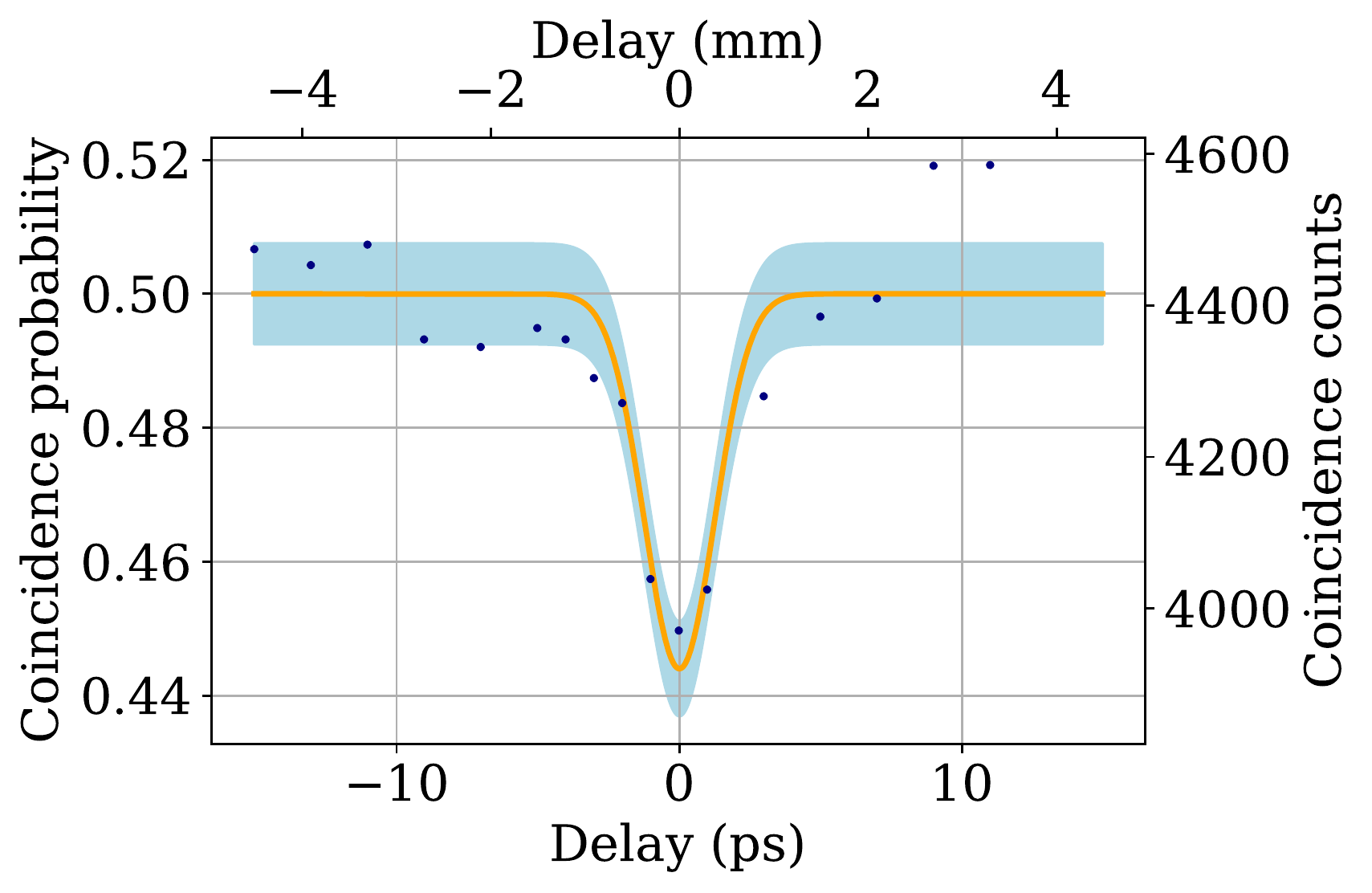}
\caption{Heralded two photon interference data. Coincidence counts are integrated over 4 hours per data point with a step size of 6~mm. The incident pump power was set to 400~mW and the parameters $\delta$ and $\sigma$ are set to match the biphoton interference measurements. The orange line is the line of best fit, the light blue region is the one sigma confidence interval assuming Poissonian counting statistics. The measured visibility is 11.2(1.4)\% which agrees with the expected value of 12.5\% for an eight-mode entangled state within experimental error.} 
\label{4folddip_plot}
\end{center}
\end{figure}

\subsection{Frequency resolved polarisation tomography} 
\label{freq pol tom}

We carry out polarisation tomography using SIC projections to limit the number of joint spectra required to 16 compared to 36 required if measuring mutually unbiased basis. Each projection contains on average $20\times10^{6}$ detected recorded over two hours of measurement. 

The SIC states used are given by,
\begin{align}
    M_1 &= \dfrac{1}{2}\left(\mathbb{I} +  \dfrac{\sigma_x + \sigma_y + \sigma_z}{\sqrt{3}}\right) \\
    M_2 &= \dfrac{1}{2}\left(\mathbb{I} +  \dfrac{\sigma_x - \sigma_y - \sigma_z}{\sqrt{3}}\right) \\
    M_3 &= \dfrac{1}{2}\left(\mathbb{I} +  \dfrac{-\sigma_x + \sigma_y- \sigma_z}{\sqrt{3}}\right) \\
    M_4 &= \dfrac{1}{2}\left(\mathbb{I} +  \dfrac{-\sigma_x - \sigma_y + \sigma_z}{\sqrt{3}}\right),
\end{align}
where $\mathbb{I}$ is the identity matrix and $\left\{\sigma_x,\sigma_y,\sigma_z\right\}$ are the Pauli matrices. 

\begin{figure*}[htb]
\centering
\label{JSI_binning}
{\includegraphics[width=0.49\columnwidth]{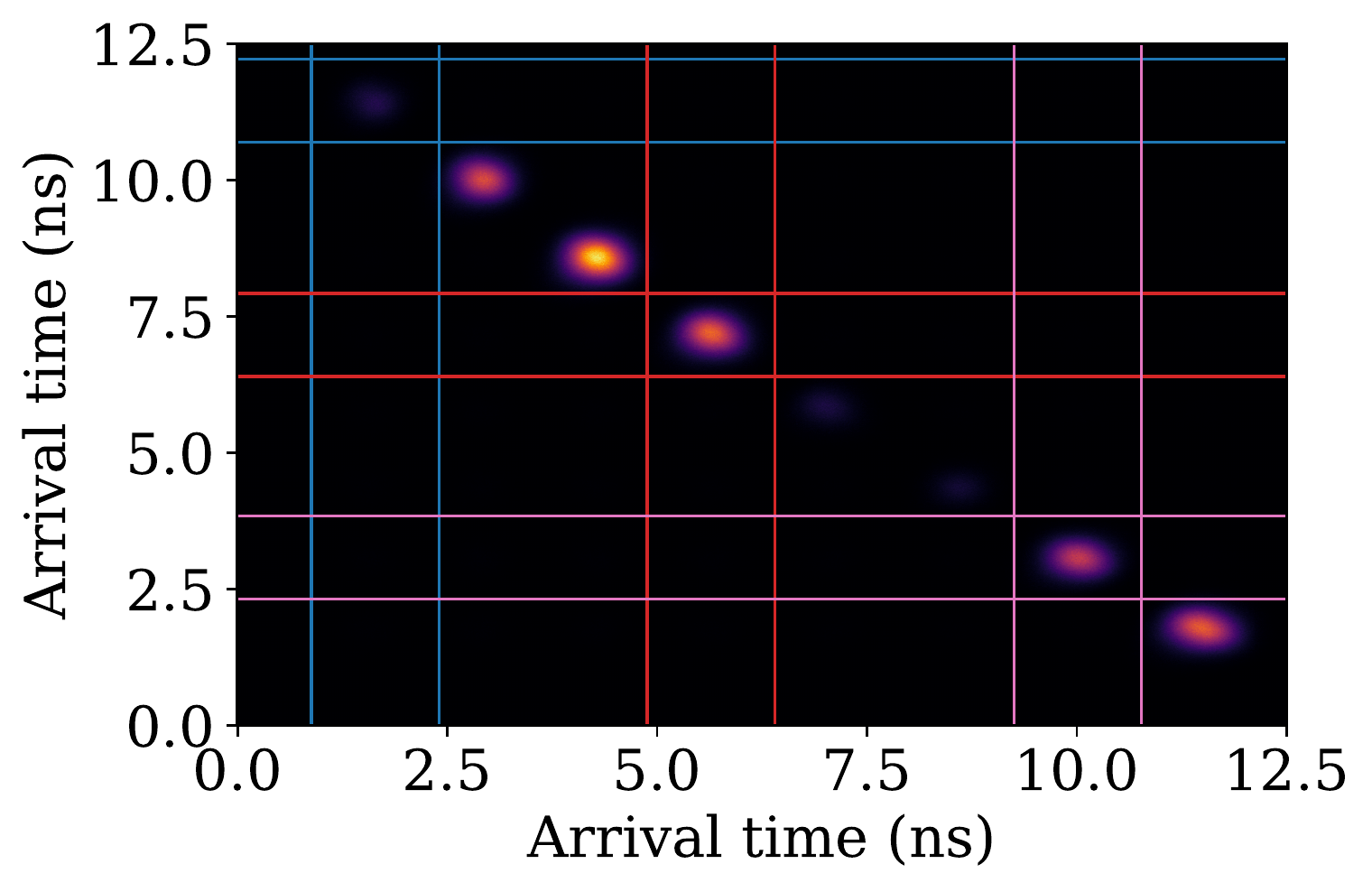}}
{\includegraphics[width=0.49\columnwidth]{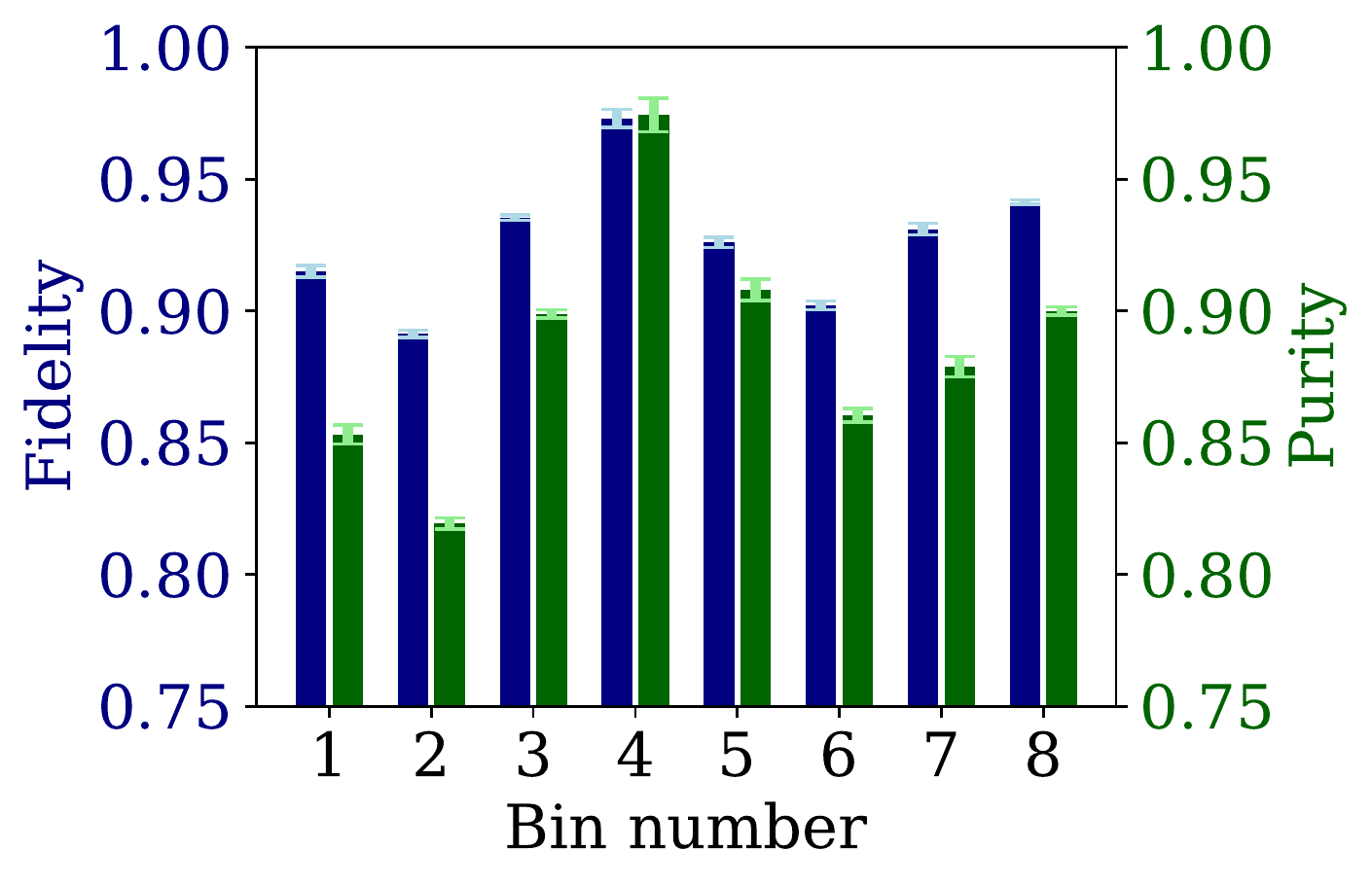}}
\caption{(Left) Experimental JSI for polarisation projection $M_1$ on both qubits.
The signal and idler filtering applied in post-processing is shown for bins 1 (blue), 4 (red) and 7 (pink). The bin width is 1.52~ns which corresponds to 3.8~nm filter width given the fibre dispersion of around 20~ps/(nm$\cdot$km) and 20 km of fiber.
(Right) Results of polarisation tomography with frequency resolved measurements. The average purity and fidelity are 88.7(3)\% and 92.6(1)\% respectively. Error bars are calculated by 1000 rounds of Monte Carlo sampling assuming Poissonian counting statistics.}
\label{polarisation TOFS}
\end{figure*}

We show the reconstructed purities and fidelities for each frequency bin after frequency resolved polarisation tomography. Example time gating used to replicate frequency filtering are shown in Fig \ref{polarisation TOFS}.

\end{document}